# A spatiotemporal oscillator model for ENSO

YaoKun Li[a]

[a] *College of Global Change and Earth System Science, Faculty of Geographical Science, Beijing Normal University, Beijing 100875, China*

*Corresponding author*: YaoKun Li, liyaokun@bnu.edu.cn

ABSTRACT

A spatiotemporal oscillator model for El Niño/Southern Oscillation (ENSO) is constructed based on the sea surface temperature (SST) and thermocline depth dynamics. The model is enclosed by introducing a proportional relationship between the gradient in SST and the oceanic zonal current and can be transformed into a standard wave equation that can be decomposed into a series of eigenmodes by cosine series expansion. Each eigenmode shows a spatial mode that oscillates with a natural frequency. The first spatial mode, that highlights SST anomaly (SSTA) contrast in the eastern and western Pacific, the basic characteristics of the eastern Pacific (EP) El Niño, oscillates with a natural period of around 4.3 years, consistent with the quasi-quadrennial (QQ) mode. The second spatial mode, that emphasizes SSTA contrast between the central and the eastern, western Pacific, the basic spatial structure of the central Pacific (CP) El Niño, oscillates with a natural period of 2.3 years that is half of the first natural period, also consistent with the quasi-biennial (QB) modes. The combinations of the first two eigenmodes with different weights can feature complex SSTA patterns with complex temporal variations. In open ocean that is far away from the coastlines, the model can predict waves propagating both eastward and westward. Besides, the net surface heating further complicates the temporal variations by exerting forced frequencies. The model unifies the temporal and spatial variations and may provide a comprehensive viewpoint for understanding the complex spatiotemporal variations of ENSO.

# 1. Introduction

The El Niño/Southern Oscillation (ENSO) phenomenon, manifested by the great swings of large-scale sea surface temperature anomalies (SSTAs) over the equatorial central to eastern Pacific oceans, is a major source of interannual global shifts in climate patterns and weather activities (Jin, 2022). ENSO originates in the tropical Pacific through interactions between the ocean and the atmosphere, but its environmental and socioeconomic impacts are felt worldwide (McPhaden et al., 2006). Thanks to the continuous observation in the tropical Pacific by the Tropical Ocean Global Atmosphere (TOGA) program, our understanding of ENSO has made significant process (Wang and Picaut, 2004; McPhaden et al., 2010), and has continued to evolve as new layers of complexity that refers to the diversity in spatial patterns, amplitude and temporal evolution (Timmermann et al., 2018).

To deal with the temporal evolution, it is natural to highlight the variation of SSTAs in the central to eastern Pacific that leads to the neglect of the west-east gradients in SSTAs, hence the spatial structure. The delayed oscillator model (Suarez and Schopf, 1988; Battisti and Hirst, 1989) introduces a time delay term to include the effects of the oceanic Rossby and Kelvin wave transit that had been noticed by McCreary (1983) in a simple coupled ocean-atmosphere model. The recharge oscillator (RO) model (Jin, 1997) combines SST dynamics and oceanic adjustment dynamics into a coupled basin-wide RO that relies on the nonequilibrium between the zonal mean equatorial thermocline depth and wind stress. There are other types of theoretical models to highlight different physical processes, such as the western Pacific oscillator model (Weisberg and Wang, 1997), the advective-reflective oscillator model (Picaut et al., 1997), and the unified oscillator model (Wang, 2001). The recent review by Wang (2018) explicitly compared the similarities and differences among these oscillator models. Specific to the RO framework, incorporating seasonality, nonlinearity, and multiscale processes, it allows for basic understanding of how key physical processes determine ENSO's properties, such as its amplitude, periodicity, phase-locking, asymmetry, and nonlinear rectification onto the mean state (Jin et al., 2020). For example, modified parameter RO models reproduces the main phase-locking characteristics found in observation and suggested that seasonal modulation of the couple coupled stability is responsible for ENSO phase locking to the annual cycle (An and Jin, 2011; Stein et al., 2014; Chen and Jin, 2020). The state-dependent stochastic forcing in RO model enhances the instability of ENSO and its ensemble spread, generates asymmetry in the predictability (Jin et al., 2007; Levine and Jin, 2010). RO framework with nonlinear advection may explain ENSO amplitude modulations and irregularity by applying the concept of homoclinic and heteroclinic connections (Jin, 1998; Timmermann and Jin, 2002; Timmermann et al., 2003).

To deal with the spatial diversity, theoretical and statistical analyses based on the observation and the model simulating data are often used to interpret the observed two types of ENSO, now widely known as central Pacific (CP) and eastern Pacific (EP) El Niño events (Larkin and Harrison, 2005a, b; Ashok et al., 2007; Kao and Yu, 2009; Kug et al., 2009). For example, linear eigen-analysis of the Zebiak-Cane (ZC) model shows that there are two leading ENSO modes that have periods of around 4 and 2 years and thereby referred to the quadrennial (QQ) and quasi-biennial (QB) modes, respectively (Bejarano and Jin, 2008; Xie and Jin, 2018). This may be further demonstrated by natural random variations in a multivariate, "patterns-based", red noise model (Newman et al., 2011). Besides, new indices, such as the Trans-Niño Index (TNI) which is given by the difference in normalized SSTAs between Niño-1+2 and Niño-4 regions (Trenberth and Stepaniak, 2001) and the E and C indices that are based on the first two empirical orthogonal function (EOF) modes of tropical Pacific SSTAs (Takahashi et al., 2011), may also be introduced to describe the diversity of patterns. The recent review by Capotondi et al. (2020) explicitly summarized the key aspects of ENSO's spatial diversity.

The previous investigations had highlighted either the temporal or the temporal variations and had greatly promoted out understanding ENSO complexity. Recent work has attempted to systematically discuss the complex spatiotemporal pattern diversity (STPD). For example, Fang and Mu (2018) extended the RO model to a three-region conceptual model to describe the entire western, central, and eastern equatorial Pacific. Takahashi et al. (2019) suggested that it is sufficient to produce the two types of ENSO in the nonlinear RO model. However, these attempts are basically extensions of the RO framework and naturally have no capacity to capture the SSTA patterns. Therefore, as Jin (2022) proposed the scientific question in a recent review paper—can we achieve a conceptual understanding of ENSO STPD through a systematic investigation of various contributing sources in a similar way to the simple RO model for our understanding of the basic dynamics of ENSO?

To answer this question and to better understanding ENSO STPD, this paper tries to provide a systematic view to integrate the spatial and temporal variations by standing on the shoulders of the previous successes. This paper is organized as below. Following this introduction section, a spatiotemporal oscillator model is established by applying proper assumptions to link the SSTA gradients with the zonal oceanic current anomaly in Section 2. The analytic solution is derived in Section 3. The solution suggests SSTA STPD can be decomposed into spatial modes associated with natural oscillation modes, each of which may be regarded as the classic RO model. The first two spatial modes and their corresponding natural oscillations, as well as the different combinations of the two modes are discussed in Section 4. The results suggest that they can reproduce the EP and CP El Niño events, with complex temporal variations. The model's assumptions and the propagation features are discussed in Section 5. Finally, a conclusion ends the paper in Section 6.

## 2. The spatiotemporal oscillator model

SSTAs for the central and eastern Pacific must be considered for capturing the essence of CP and EP pattern (Jin, 2022). However, if just introducing two variables to represent the SSTAs in the central and the eastern Pacific (e.g., $T_C$ and $T_E$), it will only follow the existing RO framework and make it more complicated. Besides, only introducing two SSTA variables are not a more precise treatment to reflect the spatial patterns of SSTAs. Therefore, to better describe the SSTA variations, let's start with the temperature equation for the upper ocean mixed layer (Deser et al., 2010)

$$\frac{\partial T}{\partial t} + u\frac{\partial T}{\partial x} + v\frac{\partial T}{\partial y} + w\frac{T-T_b}{H} = \frac{1}{\rho C_p H}Q, \qquad (1)$$

where $T$ is the mixed layer temperature (equal to the SST), $u, v$ are the zonal, meridional currents in the mixed layer, respectively, $w$ is the vertical velocity in the mixed layer, $T_b$ is the temperature of the water at depth that is entrained into the mixed layer, $H = 50$ m (Jin et al., 2020) is the mixed layer depth, $\rho$ is the density of the seawater, $C_p$ is the heat capacity, and $Q$ is the net surface heat flux. Note that for the convenience of the theoretical analysis, the horizontal current velocity is not divided into geostrophic current velocity and Ekman current velocity and the vertical velocity is not divided into vertical entrainment rate and Ekman pumping velocity as the common practice.

The SSTA equation for the equatorial band mean (e.g., 5ºS-5ºN), linearized with respect to a time-mean and area averaged basic state, then may be written as

$$\frac{\partial T'}{\partial t} + \bar{u}\frac{\partial T'}{\partial x} + K_h h' = \frac{1}{\rho C_p H} Q' \triangleq Q', \tag{2}$$

where $T'$ is the SSTA, $\bar{u} = -0.1$ m s$^{-1}$ is the zonal current of the basic state in the mixed layer and is set to a constant value, $Q' = Q - \bar{Q}$ is the net radiation anomaly with respect to its basic state $\bar{Q}$. Note that the fourth term in Eq. (1), representing the vertical advection, is not linearized but expressed as $K_h h'$ where $K_h$ is a coefficient and $h'$ is the thermocline depth anomaly (TDA) according to Battisti and Hirst (1989). The value for $K_h$ may vary from $3.5 \times 10^{-9}$ K m$^{-1}$ s$^{-1}$ (Hirst, 1986, 1988) to $1.7 \times 10^{-8}$ K m$^{-1}$ s$^{-1}$ (Kang and An, 1998) and may also vary with $y$ (e.g., as Kang and An (1998) and Kang et al. (2001) suggested, it has a large value near the equator and a small value off the equator). Considering the two values, a moderate value, namely, $K_h = 1.0 \times 10^{-8}$ K m$^{-1}$ s$^{-1}$ is specified in this investigation. Eq. (2) may become a simpler form if replacing the net surface heat flux with a Newtonian cooling $-\alpha T'$ where $\alpha > 0$ is the damping coefficient.

The TDA is one of the main variables that needs to be considered for understanding ENSO dynamics (Jin, 1997). The linearized TDA equation with respect to the zonal basic current and the undisturbed mixed layer depth is

$$\frac{\partial h'}{\partial t} + \bar{u}\frac{\partial h'}{\partial x} + H\frac{\partial u'}{\partial x} = 0. \tag{3}$$

It contains the advection by the zonal basic current and the divergence of the zonal current anomalies.

Eqs. (2) and (3) constitute the system that describes the main dynamics for understanding ENSO. However, it is not enclosed since an additional variable, the zonal current anomaly, is introduced. Although it may be enclosed by introducing the zonal momentum equation, it will make the system more complicated and will not be conducive to analytical analysis. Before enclosing the system by introducing approximation relations, let's revisit the practice in constructing the RO model (Jin, 1997), that is,

$$\tau' = b_1 T', \tag{4}$$

where $\tau'$ is the zonal wind stress anomaly, $b_1$ is coupling coefficient. This simple approximation relation of wind stress anomaly and SSTA works since the atmospheric response to a warm SSTA of the central to eastern Pacific is a westerly and the SSTA variable in RO model is averaged over the central to eastern Pacific (Jin, 1997). However, this approximation does not work if the SSTA variable is not limited in the central to eastern Pacific as Eq. (2) describes.

According to Gill (1980), the atmospheric response to diabatic heating (which is equivalent to SSTA as Neelin (1989) had pointed out) is a westerly wind west of the heating while an easterly wind east of the heating. Actually, Jin (1997) also realized this. He still proposed the approximation relation because the SSTA variable is averaged over the central to eastern Pacific so that the westerly anomaly west of the warm SSTA is the main consideration and because there is an overall westerly (easterly) anomaly for a positive (negative) SSTA over the entire basin of the equatorial band (e.g., Fig. 1a). To capture the essence of the atmospheric response to a warm SSTA—westerly (easterly) wind stress anomaly west (east) of a warm SSTA, a simple approximation relation is proposed, that is,

$$\tau' = b_2 \frac{\partial T'}{\partial x}, \tag{5}$$

where $b_2$ is the coupling coefficient. This relation links the wind stress anomaly with the gradient of SSTA. As shown in Fig. 1b, the performance of this approximation relation is better than that the approximation relation in Eq. (4). Considering the fact that the oceanic current in the mixed layer in general follows the direction of the wind stress (e.g., see Fig. 1c), a simple proportional relation between then may be established, namely,

$$u' = b_3 \tau', \tag{6}$$

where $b_3$ is the coefficient.

The above two approximation relations build a linkage between the zonal current anomaly and the gradient of SSTA, that is,

$$u' = b \frac{\partial T'}{\partial x}, \tag{7}$$

where $b = 1.25 \times 10^5$ m$^2$ s$^{-1}$ K$^{-1}$ is the coupling coefficient. Its value is determined by linear regression the gradient of the SSTA data to the oceanic current data in the simple ocean data assimilation (SODA) version 3.4.2 (Carton et al., 2018). The correlation coefficient between them is 0.18. Although the correlation coefficient looks low, it passes the 1% significance level due to large size samples (larger than 20,000 samples).

Substituting Eq. (7) into Eq. (3) to eliminate $u'$, a linear coupled system with both thermocline dynamics and SSTA dynamics may be derived

$$\mathcal{L}T' + K_h h' = Q',  \quad (8)$$

$$\mathcal{L}h' + Hb\frac{\partial^2 T'}{\partial x^2} = 0, \quad (9)$$

where $\mathcal{L} = \frac{\partial}{\partial t} + \bar{u}\frac{\partial}{\partial x}$. Note that the system may be seen as the simplified version of the coupled air sea interaction models, such as Zebiek and Cane (ZC) model (Cane and Zebiak, 1985; Cane et al., 1986; Zebiak and Cane, 1987). On the other hand, the system contains two variables: SSTA and TDA, the same as the RO model but with spatial variations. Therefore, it may also be seen as a development of the classic RO model.

Eliminating $h'$ from the system, a wave equation may be derived as

$$\mathcal{L}^2 T' = c^2 \frac{\partial^2 T'}{\partial x^2} + \mathcal{L}Q' \quad (10)$$

where $c = \sqrt{K_h Hb} = 0.25$ m s$^{-1}$ is the propagation speed of the wave. Introducing the coordinate system that moves with the basic zonal current

$$x_1 = x - \bar{u}t, t_1 = t, \quad (11)$$

Eq. (10) may be further reduced to

$$\frac{\partial^2 T'}{\partial t_1^2} = c^2 \frac{\partial^2 T'}{\partial x_1^2} + \frac{\partial Q'}{\partial t_1}. \quad (12)$$

Eq. (12) is a standard form wave equation and is easy to solve.

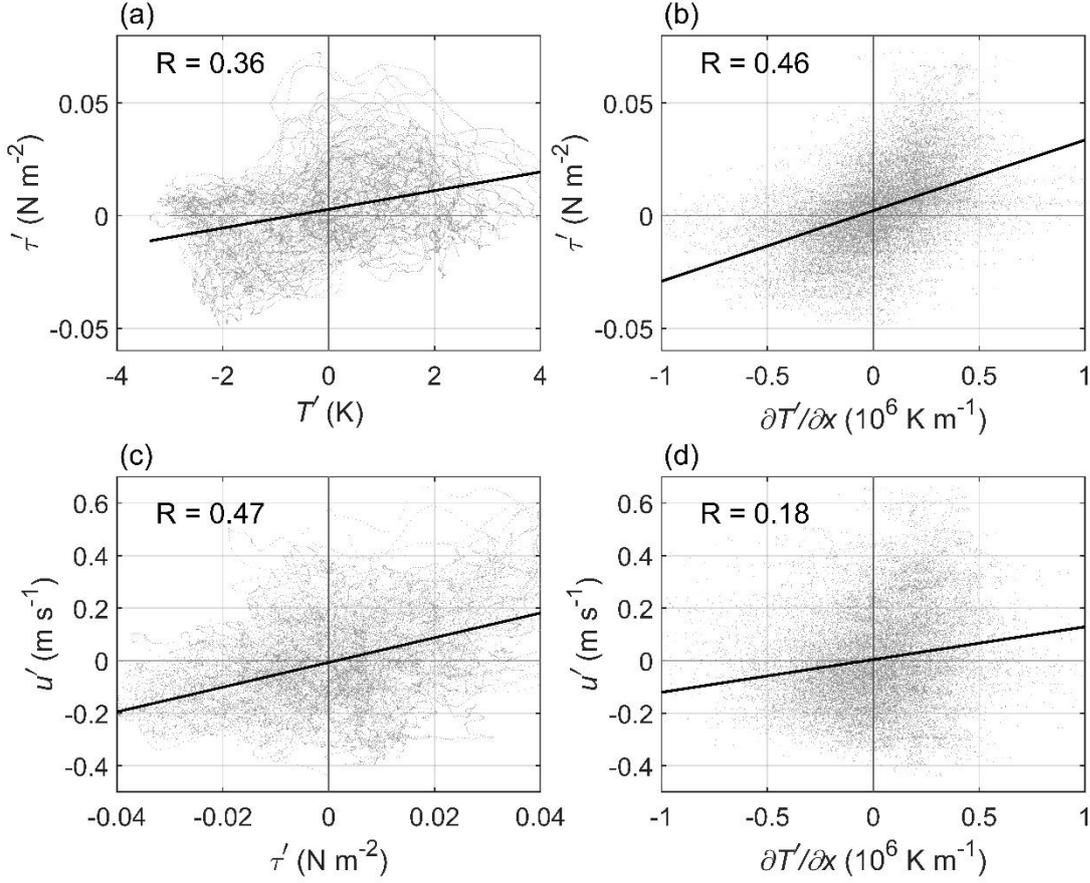

Fig. 1 The linear relations between the SSTA and the zonal wind stress anomaly (a), between the gradient of the SSTA and the zonal wind stress anomaly (b), between the zonal wind stress anomaly and the oceanic current velocity anomaly (c), and between the gradient of the SSTA and the oceanic current velocity anomaly (d). The data comes from the simple ocean data assimilation (SODA) version 3.4.2.

## 3. The solution

To make the deriving process simpler and clearer, the solution for Eq. (12) is derived and presented below. After the solution is derived, it is easy to write the solution for Eq. (10) according to the coordinate transform Eq. (11).

The initial values are specified as

$$T'(x_1)\big|_{t_1=0} \equiv F(x_1), \tag{13}$$

$$h'(x_1)\big|_{t_1=0} \equiv G(x_1), \tag{14}$$

where $F(x_1)$ and $G(x_1)$ are known functions. The well-posed free boundary conditions are specified as

$$\left.\frac{\partial T'}{\partial x_1}\right|_{x_1=0} = 0, \left.\frac{\partial T'}{\partial x_1}\right|_{x_1=L} = 0, \tag{15}$$

where $L$ is the basin width of the equatorial Pacific. Expanding $T'$ and $h'$ into cosine series, e.g.,

$$\{T', Q'\} = \sum_{n=0}^{\infty} \{T_n(t_1), Q_n(t_1)\} \cos(\lambda_n x_1), \tag{16}$$

where $\lambda_n = \dfrac{n\pi}{L}$ will naturally satisfy the boundary conditions Eq. (15). Similarly, $F(x_1)$ and $G(x_1)$ are also expanded as cosine series

$$\begin{bmatrix} F(x_1) \\ G(x_1) \end{bmatrix} \equiv \begin{bmatrix} T'(x_1) \\ h'(x_1) \end{bmatrix}\bigg|_{t_1=0} = \sum_{n=0}^{\infty} \begin{bmatrix} T_n(0) \\ h_n(0) \end{bmatrix} \cos(\lambda_n x_1) \equiv \sum_{n=0}^{\infty} \begin{bmatrix} f_n \\ g_n \end{bmatrix} \cos(\lambda_n x_1). \tag{17}$$

It is easy to derive a second order differential equation with constant coefficients by substituting these cosine series into Eq. (12), that is,

$$\frac{d^2 T_n}{dt_1^2} + \omega_n^2 T_n = \frac{dQ_n}{dt_1}, \tag{18}$$

where $\omega_n = c\lambda_n$, $n = 1, 2, 3, \cdots$ is integer. The solution for Eq. (18) is

$$T_n(t_1) = A_n \cos(\omega_n t_1) + B_n \sin(\omega_n t_1) + \frac{1}{\omega_n} \int_0^{t_1} \frac{dQ_n}{d\tau} \sin\omega_n(t_1 - \tau) d\tau, \tag{19}$$

where $A_n = f_n$, $B_n = \dfrac{q_n - K_h g_n}{\omega_n}$, $q_n = Q_n(0)$ are coefficients that are determined by initial values. Eq. (18) indicates a forced oscillation. Its free oscillation part is formally the same as the harmonic oscillation in the original RO model (Jin, 1997). Besides, if transforming to the coordinate system $(x, t)$, Eq. (18) will be a damped or excited oscillation, formally the same as the RO model. It is obvious that Eq. (18) further extends the RO model to associate with spatial modes. Or, the original RO model only reflects the oscillation of a specified spatial mode. In this sense, it may be regarded as a spatial RO model.

For $n=0$, Eq. (18) becomes

$$\frac{d^2 T_0}{dt_1^2} = \frac{dQ_0}{dt_1}, \tag{20}$$

and the solution is

$$T_0(t_1) = A_0 + B_0 t_1 + \int_0^{t_1} Q_0(\tau) d\tau, \tag{21}$$

where $A_0 = f_0$, $B_0 = 0$ are coefficients that are determined by initial conditions. Note that $B_0$ should be zero to make sure SSTA is limited when time is long enough. $A_0$ is a constant value and equals to the constant term in cosine expansion of SSTA. $\int_0^{t_1} Q_0(\tau) d\tau$ is a time related term that relies on the integral of the constant term in cosine series of the net surface heating.

The solution for Eq. (12) can be eventually expresses as

$$\begin{aligned} T'(x_1, t_1) = f_0 &+ \int_0^{t_1} Q_0(\tau) d\tau \\ &+ \sum_{n=1}^{\infty} \left[ f_n \cos(\omega_n t_1) + \frac{q_n - K_h g_n}{\omega_n} \sin(\omega_n t_1) \right] \cos(\lambda_n x_1), \\ &+ \sum_{n=1}^{\infty} \frac{1}{\omega_n} \left[ \int_0^{t_1} \frac{dQ_n}{d\tau} \sin \omega_n (t_1 - \tau) d\tau \right] \cos(\lambda_n x_1) \end{aligned} \tag{22}$$

or

$$\begin{aligned} T'(x_1, t_1) = f_0 &+ \int_0^{t_1} Q_0(\tau) d\tau + \sum_{n=1}^{\infty} C_n \sin(\omega_n t_1 + \theta_n) \cos(\lambda_n x_1) \\ &+ \sum_{n=1}^{\infty} \frac{1}{\omega_n} \left[ \int_0^{t_1} \frac{dQ_n}{d\tau} \sin \omega_n (t_1 - \tau) d\tau \right] \cos(\lambda_n x_1) \end{aligned}, \tag{23}$$

where $C_n = \sqrt{f_n^2 + \left(\frac{q_n - K_h g_n}{\omega_n}\right)^2}$ is the amplitude, $\theta_n$ is the phase and $\sin \theta_n = \frac{f_n}{C_n}$, $\cos \theta_n = \frac{q_n - K_h g_n}{C_n \omega_n}$. Except for the constant term that is determined by the initial SSTA pattern, and the time related term that hat represent the influence of the initial net surface heating, Eq. (22) or Eq. (23) demonstrates that the spatiotemporal variation of SSTA can be decomposed into the summation of a series of natural oscillation with a frequency of $\omega_n$ and a series of forced oscillation that are associated with the corresponding spatial modes.

For a specific spatial mode (ignoring the net surface heating), each point of the spatial mode oscillates with its natural frequency and with the same phase but varying amplitude of $|C_n \cos \lambda_n x_1|$ that depends on its location in the spatial mode. This can also be called a stationary wave of zero phase speed. In other words, this specific spatial mode ($\cos \lambda_n x_1$) oscillates with a natural frequency $\omega_n$, a phase $\theta_n$, and an amplitude $C_n$. The temporal variation will become more complex when considering the effect of the net surface heating forcing which will add at least one external forced frequency to the natural frequency for a specific spatial mode. Specially, if a forced frequency is just right equal to the natural frequency, the corresponding amplitude will tend to be infinity due to resonance. Specific calculation cases will be given in the next section. Besides, although the solution denotes oscillation with no propagation, there should be certain propagation feature when an initial coastal SSTA perturbation whose propagation direction is far away from the other coastline due to large basin width of the equatorial Pacific. This will be discussed later.

Transforming to the $(x,t)$ coordinate system, Eq. (23) becomes

$$T'(x,t) = f_0 + \int_0^t Q_0(\tau)d\tau + \sum_{n=1}^{\infty} C_n \sin(\omega_n t + \theta_n)\cos \lambda_n (x - \bar{u}t) \\ + \sum_{n=1}^{\infty} \frac{1}{\omega_n}\left[\int_0^t \frac{dQ_n}{d\tau}\sin \omega_n (t-\tau)d\tau\right]\cos \lambda_n (x - \bar{u}t) \tag{24}$$

Since $\cos \lambda_n (x - \bar{u}t) = \cos \lambda_n x \cos \bar{u}t + \sin \lambda_n x \sin \bar{u}t$, Eq. (24) becomes

$$T'(x,t) = f_0 + \int_0^t Q_0(\tau)d\tau \\ + \sum_{n=1}^{\infty} C_n \left[\sin(\omega_n t + \theta_n)\cos \bar{u}t\right]\cos \lambda_n x \\ + \sum_{n=1}^{\infty} \frac{1}{\omega_n}\left[\cos \bar{u}t \int_0^t \frac{dQ_n}{d\tau}\sin \omega_n (t-\tau)d\tau\right]\cos \lambda_n x \\ + \sum_{n=1}^{\infty} C_n \left[\sin(\omega_n t + \theta_n)\sin \bar{u}t\right]\sin \lambda_n x \\ + \sum_{n=1}^{\infty} \frac{1}{\omega_n}\left[\sin \bar{u}t \int_0^t \frac{dQ_n}{d\tau}\sin \omega_n (t-\tau)d\tau\right]\sin \lambda_n x \tag{25}$$

The last two terms in Eq. (25) represent temporal variations associating with spatial modes of sine curves. It is obvious that the different locations in the basin will have different phases due to modulation of the sine and cosine spatial modes. In other words, Eq. (25) may manifest travelling waves. Eq. (25) is equivalent to the solution of Cauchy problem for Eq. (10) with no specified boundary conditions. It means that SSTA perturbations may propagate across the coastlines as if the coastlines do not exist. However, considering the fact that waves may be reflected or absorbed due to the blocking effect of the coastlines, Eq. (25) may be applied to analyze the wave propagation toward a far distant coastline so that the influence of the boundary can be ignored. This limitation comes from the practice that specifies a basin-wide constant zonal basic current that should have been zero at the coastlines. The constant zonal basic current over the entire basin means that it is hard to specify well-posed boundary conditions at the coastlines. Therefore, a spatial varying zonal basic current that vanishes at the coastlines seems a better choice. Besides, an external frequency that associates with the zonal basic current is added on the temporal oscillations. Generally speaking, the zonal basic current further complicates the solution in both spatial and temporal dimensions.

## 4. Results

To provide a fundamental scene for understanding the spatiotemporal oscillator model, the results are provided based on Eq. (23). It is equivalent to moving with the zonal basic current or ignoring the basic zonal current for easy understanding. The first spatial mode ($n=1$) is a standard cosine curve in the $[0, L]$ range with a period of $2\pi$, hence a SSTA pattern with warm (or cold) SSTA in the western Pacific but cold (or warm) SSTA in the eastern Pacific. This spatial distribution is quite similar to the SSTA pattern in the EP El Niño events. The second spatial mode ($n=2$) is a standard cosine in the $[0, L]$ range with a period of $\pi$, hence a SSTA pattern with cold (or warm) SSTA in the central Pacific but warm (or cold) SSTA in the eastern and western Pacific. This spatial distribution is quite similar to the SSTA pattern in the CP El Niño events. Note that even if the first two modes are similar to the EP and CP El Niño events, they are not the observed EP and CP El Niño events. Actually, an EP or CP El Niño event should be seen as the combination of the first two modes with different weights. For higher order modes, SSTA patterns associate with small scale structures and do not have no clear physical image so far. Therefore, only the first two spatial modes can represent the basic SSTA pattern in ENSO events.

*a. Natural Oscillation*

To highlight the first two natural spatiotemporal oscillation modes, their expressions are written separately

$$T_n = \sin\left(\omega_n t_1 + \frac{\pi}{2}\right) \cos(\lambda_n x_1), \qquad (26)$$

where $n = 1, 2$. The initial phases are set to be $\frac{\pi}{2}$, denoting that the SSTA in each mode has maximum value while the TDA equals zero at the beginning time. They are also summed with different weight to highlight their relative importance in constructing SSTA patterns, that is

$$T' = 2T_1 + T_2, \qquad (27)$$

and

$$T' = T_1 + 2T_2. \qquad (28)$$

For the first mode ($n = 1$), SSTA in the eastern and western Pacific (Fig. 2a) oscillates with alternative warm and cold anomalies and with a natural period of $T_1 = \frac{2L}{c} \approx 4.6$ years. For the second mode ($n = 2$), SSTA (Fig. 2b), with warm anomalies appearing in the central Pacific while cold anomalies in the western and eastern Pacific, oscillates with a natural period that is half of the first mode, that is, $T_2 = \frac{L}{c} \approx 2.3$ years. It is interesting to point out that these two natural periods are consistent with the previous researches that suggested the EP and CP El Niño events oscillate with QQ and QB modes (Ren and Jin, 2013; Xie and Jin, 2018). If the first mode is more dominant, the warm SSTA in the eastern Pacific can extend from the 80ºW to around 180º at the beginning time (Fig. 2c), larger than the warm SSTA in the first mode where the warm and cold anomalies halve the equatorial Pacific. On the other hand, the cold SSTA in the eastern Pacific shrinks to 140ºW-80ºW range. The SSTA pattern in this combination looks quite similar to the SSTA pattern in the strong El Niño events, such as in 1982/83, 1997/98, and 2015/16, in which warm SSTAs appear in the central to eastern Pacific. It implies that these strong El Niño events may rely on the coordination between the two modes. On the other hand, if the second mode is more dominant, the warm SSTA still locates in the central Pacific at the beginning time (Fig. 2d), that is quite similar to case for the second mode alone (Fig. 2b). The variation in the spatial pattern shows the regulation effect of the first mode is relatively minor in the CP El Niño events.

The combinations of the first two modes mean that SSTA patterns oscillate with periods of both the first and the second modes. The time series of NINO3.4 index (Fig. 3) is further portrayed to demonstrate the temporal variations due to the superposition of the two natural periods. For the cast that the first spatial mode is more dominant as Eq. (27) exhibits, NINO3.4 index (Fig. 3a) decreases from a maximum value of around 1.2K to -0.7K within around 13 months and then increases to an extreme value of around 0.1K within around 27 months to finish a cycle, the natural period of the second mode. Within the next 27 months, it moves to the same minimum value and then the maximum value to form a larger period of 54 months, the natural period of the first mode. If an El Niño event is defined by NINO3.4 index that is larger than 0.5K while La Niña event by the index smaller than -0.5K, it is quite interesting to find that two La Niña events will follow an El Niño event. For the case that the second spatial mode is more dominant as Eq. (28) shows, NINO3.4 index (Fig. 3b) has the similar variation trend as the case that the first mode is more dominant but with different amplitude. NINO3.4 index is larger than 0.5K (or around 1K) near the 27 months so that an El Niño event can be defined, which blocks the two La Niña events in a row as they occurred successively in the previous case. Therefore, El Niño and La Niña events happen alternatively in this case. Besides, the intensity of two adjacent El Niño events is also different. It is obvious that the situation will be more complex if considering different combination weights, different initial phases. This is right the capacity of the model for understanding ENSO STPD.

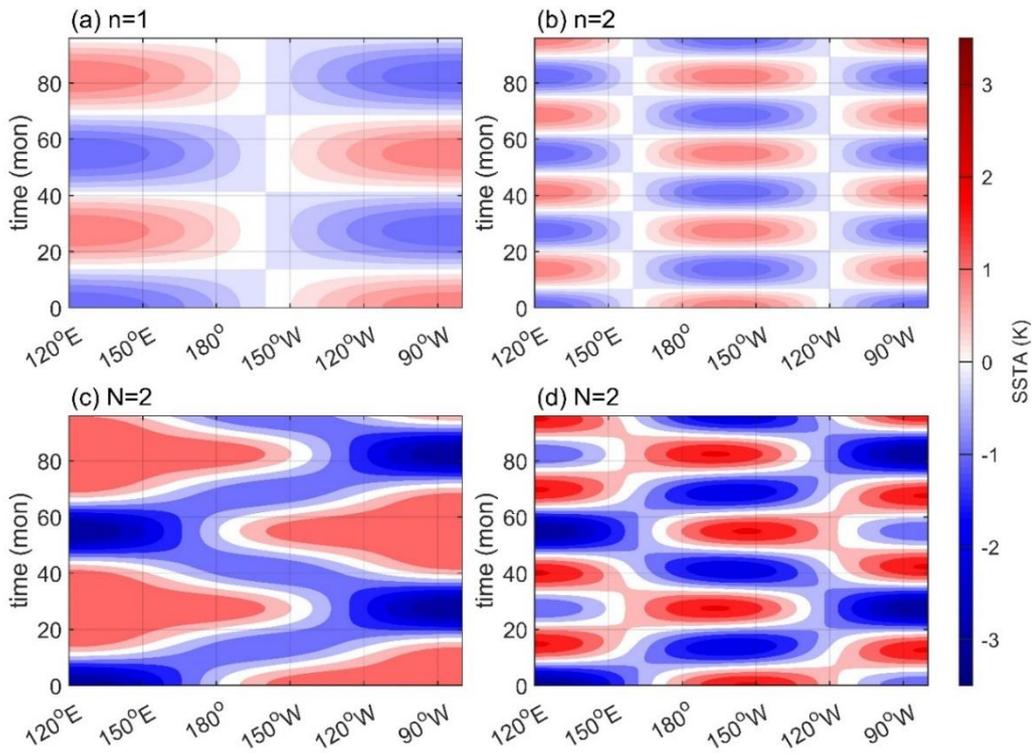

Fig. 2 The temporal evolution for the first mode (a), the second mode (b), the combination of them as Eq. (27) declares (c), and as Eq. (28) declares (d).

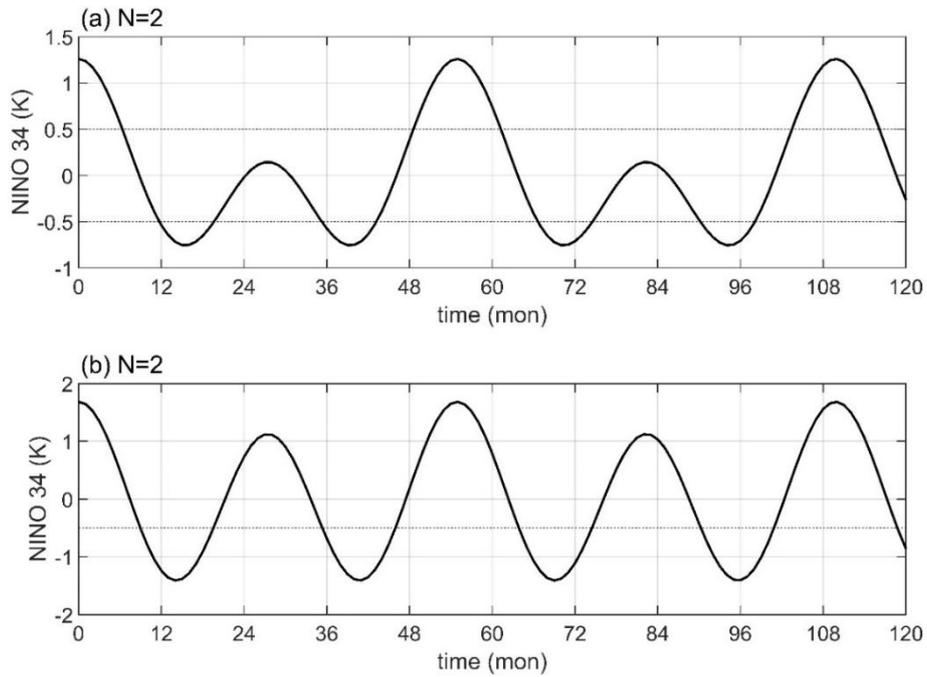

Fig. 3 The variations in NINO3.4 index for the combinations of the two modes as shown in Fig. 2c, d.

*b. Forced Oscillation*

To consider the effect of the net surface heat flux forcing, cosine expansion is conducted for the 5ºS-5ºN mean equatorial Pacific net surface heat flux anomaly in the SODA version 3.4.2 data. Spectrum analysis suggests that the first spatial mode has a most significant oscillation period of 12 months (Fig. 4a) while the second spatial mode has a significant oscillation period of around 114 months (Fig. 4b). Therefore, these two significant periods are added to two forcing modes, that are analytically expressed as

$$Q_n = \bar{Q}_0 \cos(\hat{\omega}_n t_1) \cos(\lambda_n x_1), \tag{29}$$

where $n = 1, 2$, $\hat{\omega}_1 = \dfrac{2\pi}{12}$ month$^{-1}$ represents the significant period of 12 months for the first spatial mode, $\hat{\omega}_2 = \dfrac{2\pi}{114}$ month$^{-1}$ denotes the significant oscillation periods of around 114 months (around 9.5 years) for the second spatial mode, $\bar{Q}_0 = 4.89 \times 10^{-8}$ K s$^{-1}$ equivalent to a relatively weak net surface heating flux of 10W m$^{-2}$. According to Eq. (23) and Eq. (26), SSTA variations for the first and second modes are

$$\begin{aligned}T_n(x_1, t_1) &= \left[\sin\left(\omega_n t_1 + \frac{\pi}{2}\right) - \frac{\hat{\omega}_n}{\omega_n} \int_0^{t_1} \sin\hat{\omega}_n \tau \sin\omega_n(t_1 - \tau) d\tau \right] \cos(\lambda_n x_1) \\ &= \left[\sin\left(\omega_n t_1 + \frac{\pi}{2}\right) - \frac{\hat{\omega}_n}{\omega_n} \frac{1}{\hat{\omega}_n^2 - \omega_n^2}(\hat{\omega}_n \sin\omega_n t_1 - \omega_n \sin\hat{\omega}_n t_1)\right] \cos(\lambda_n x_1)\end{aligned}. \tag{30}$$

The forced oscillations (Fig. 5) are basically the same as the previous natural oscillations (Fig. 2). For the first mode ($n=1$), except for the significant period of around 4.6 years, the SSTA (Fig. 5a) also oscillates with the forced period of 1 year even though its modulation of the SSTA is relatively minor due to relative weak intensity of the net surface heating. For the second mode ($n=2$), the situation is similar. It also oscillates with a natural period and a forced period. Similar to the natural oscillation subsection, the two modes in Eq. (30) are also be combined with different weights as Eq. (27) and Eq. (28) suggest. For the case that the first mode is more dominant, the warm SSTA pattern (Fig. 5c) also looks similar to the EP El Niño events but with significant difference with the previous natural case. For example, the warm SSTA has a largest spatial range at the beginning. It gradually shrinks to the eastern Pacific but with strengthening intensity. After the SSTA arrives the peak, it will quickly cool to enter a strong La Niña event, which will have the matched intensity and spatial range as the El Niño event. The time series of NINO3.4 index (Fig. 6a) oscillates with two significant periods of 4.6 and 2.3 years—the natural periods of the first two modes. Although the forced period is not significant and cannot be seen directly in Fig .6a, it indeed exhibits its influence on complicating the time series to make every short period cycle is slightly different from the other in a longer period. For the case that the second mode is more dominant, the variations in SSTA pattern (Fig. 5d) are quite similar to that in the previous natural case. Modulated by the natural periods and the forced period, the time of NINO3.4 index (Fig. 6b) also become more complex. It is easy to summarize that the net surface heating will further complicate ENSO STPD by inserting forced periods. Note that the first and second spatial modes for the net surface heating also have significant periods of around 60 and 30 months (see Fig. 4) that are close to the corresponding natural periods. And if these two forced periods are specified in Eq. (29), they will induce resonance to produce unreasonable strong NINO3.4 index oscillation as Fig. 7 exhibits. During a time period, a forced frequency may be close to a natural frequency. This may breed significant SSTA variations that may be provide a plausible explanation for strong ENSO events in this time period.

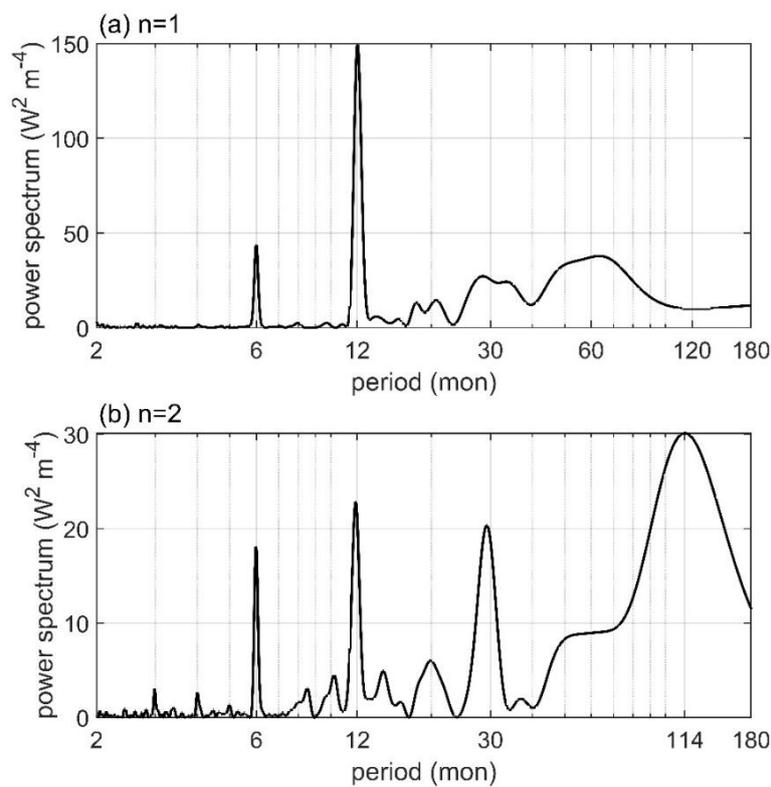

Fig. 4 The power spectrum of the time series for the first two spatial modes.

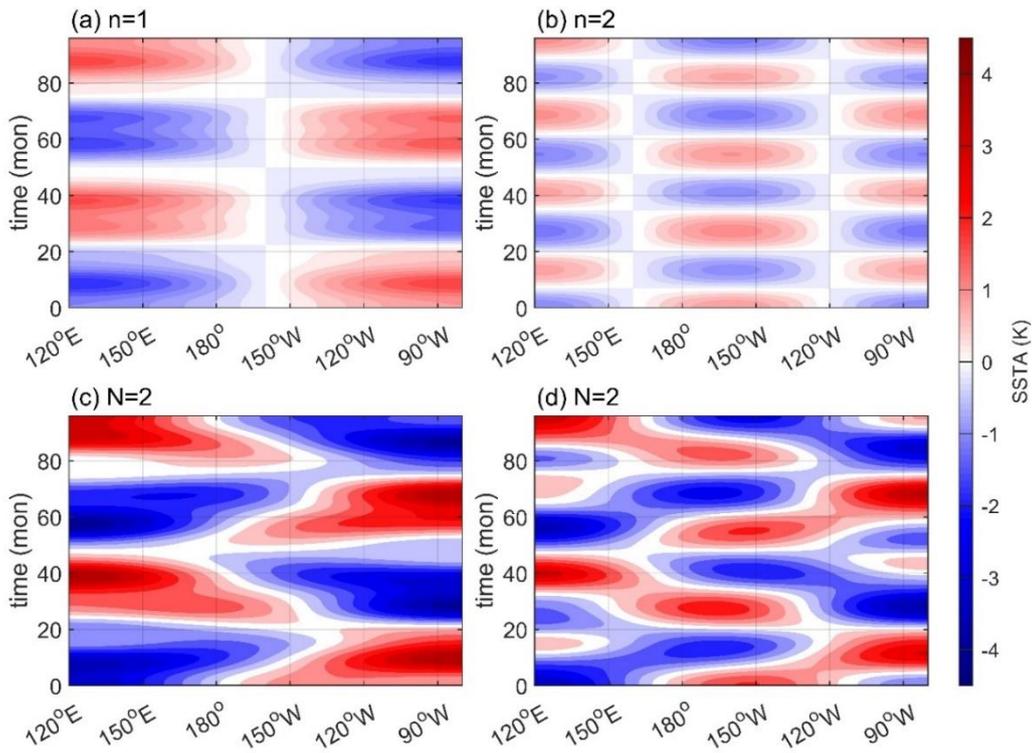

Fig. 5 The same as Fig. 2, but for the forced oscillation.

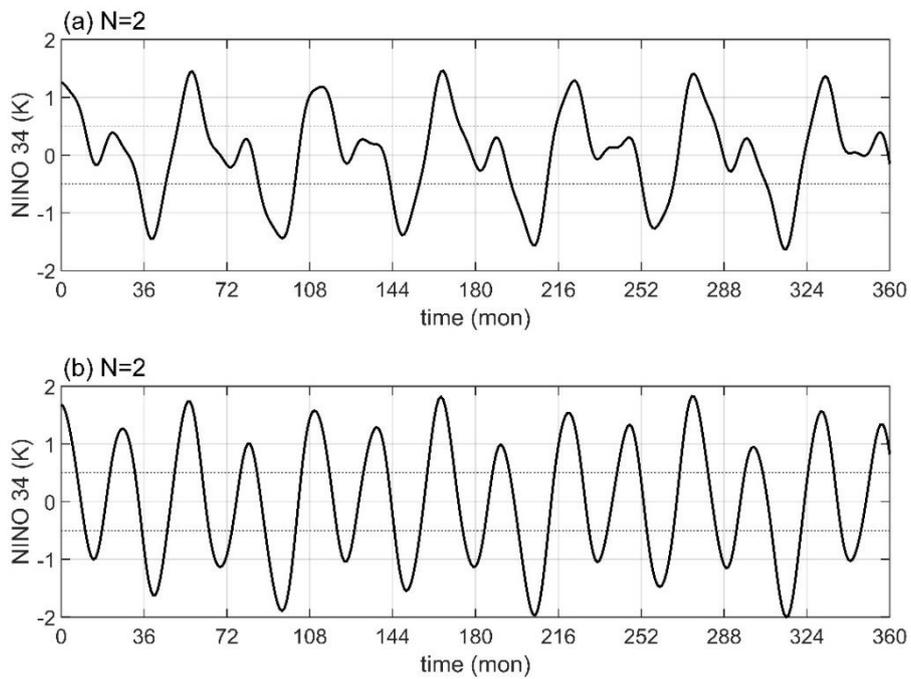

Fig. 6 The same as Fig. 3, but for the forced oscillation.

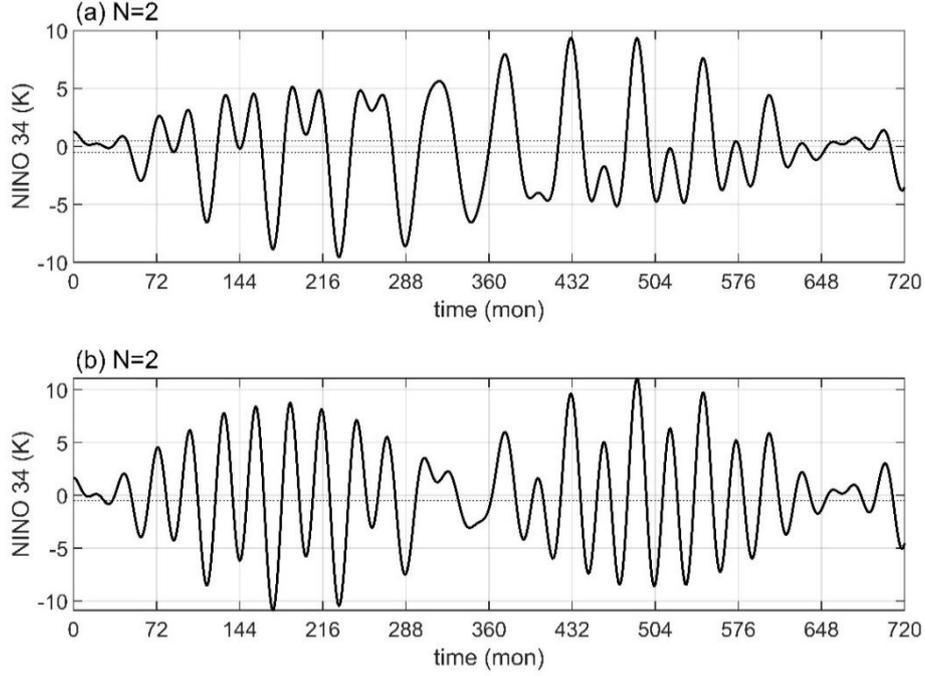

Fig. 7 The same as Fig. 6, but the forced periods are set to 60 and 30 months for the first and second modes, respectively.

## 5. Discussions

*a. Relation Between the SSTAs and the Wind Stress Anomalies*

Eq. (5) characterizes the symmetric westerly and eastly wind stress anomalies on the west and east side of a warm SSTA. However, this simple relation may be a little rough to capture the relatively weak easterly anomalies on the east side of the warm SSTA. Since the wind stress anomalies are closely linked with the SSTA and its gradient (see Fig. 1a, b), an improved simple relation may be proposed by combing Eq. (4) and Eq. (5), that is

$$\tau' = b_3 T' + b_4 \frac{\partial T'}{\partial x}, \tag{31}$$

where $b_3$ and $b_4$ are coupling coefficients. This relation may also be found in the classic paper of Gill (1980), in which the zonal wind stress anomaly is analytically expressed as a function of the heating and its gradient (e.g., see Eqs. (4.2) and (4.3) in his paper). Eq. (31) can improve the statistical relationship among SSTA, the gradient of SSTA, and the zonal wind stress anomaly (Fig. 8a) and increase the correlation coefficient to 0.54. Based on this relation, Eq. (7) becomes

$$u' = b_5 T' + b_6 \frac{\partial T'}{\partial x}, \tag{32}$$

where $b_5 = 0.0446 \text{ m s}^{-1} \text{ K}^{-1}$ and $b_6 = 8.1137 \times 10^4 \text{ m}^2 \text{ s}^{-1} \text{ K}^{-1}$ are coupling coefficients that are determined from the SODA version 3.4.2 data. Eq. (32) significantly improve the relation among SSTA, the gradient of SSTA, and the surface oceanic current anomaly (Fig. 8b). The value of the correlation coefficient is as high as 0.42, far greater than the previous 0.18.

With this improved relation, Eq. (12) becomes

$$\frac{\partial^2 T'}{\partial t_1^2} = a \frac{\partial T'}{\partial x_1} + c^2 \frac{\partial^2 T'}{\partial x_1^2} + \frac{\partial Q'}{\partial t_1}, \tag{33}$$

where $a = K_h H b_5 \approx 2.23 \times 10^{-8} \text{ m s}^{-2}$ and $c = \sqrt{K_h H b_6} \approx 0.20 \text{ m s}^{-1}$ have dimensions of acceleration and velocity respectively. It is easy to introduce $T' = \hat{T} e^{-\sigma x_1}$ and $Q' = \hat{Q} e^{-\sigma x_1}$ where $\sigma = \frac{a}{2c^2}$ to reduce Eq. (33) to a standard form

$$\frac{\partial^2 \hat{T}}{\partial t_1^2} = c^2 \frac{\partial^2 \hat{T}}{\partial x_1^2} - \mu^2 \hat{T} + \frac{\partial \hat{Q}}{\partial t_1}, \tag{34}$$

where $\mu^2 = \frac{a^2}{4c^2}$. It has the same form as Eq. (12) except for an additional $-\mu^2 \hat{T}$ term and it is also easy to solve by expanding $\hat{T}$ to cosine series as previous. Substituting the series expansion solution into Eq. (34), we derive

$$\frac{d^2 T_n}{dt_1^2} + \varpi_n^2 T_n = \frac{dQ_n}{dt_1}, \tag{35}$$

Where $\varpi_n^2 = \omega_n^2 + \mu^2$, $n = 0, 1, 2, \cdots$. It has the same form as Eq. (18) but with modified natural frequencies. Therefore, the influence of the new introduced parameter will be discussed, rather than the solution in detail. It is easy to note that the natural frequency will be modified by the parameter $\mu$ and hence $a$ to become no longer simple multiple relation. For example, the first natural period ($n = 1$) becomes to around 3 years while the second natural period ($n = 2$) is around 2.2 years. Besides, there will exist a new zeroth order ($n = 0$) natural period of around 3.6 years (the corresponding natural frequency equals $\mu$), that is close to the first order natural period. In addition to variations in temporal oscillations, the spatial modes are also modified by the factor $e^{-\sigma x_1}$ that will violate the cosine spatial modes. Nonetheless, the theoretical framework is still the same and no explicit explanation of this solution is provided and this may be further investigated later.

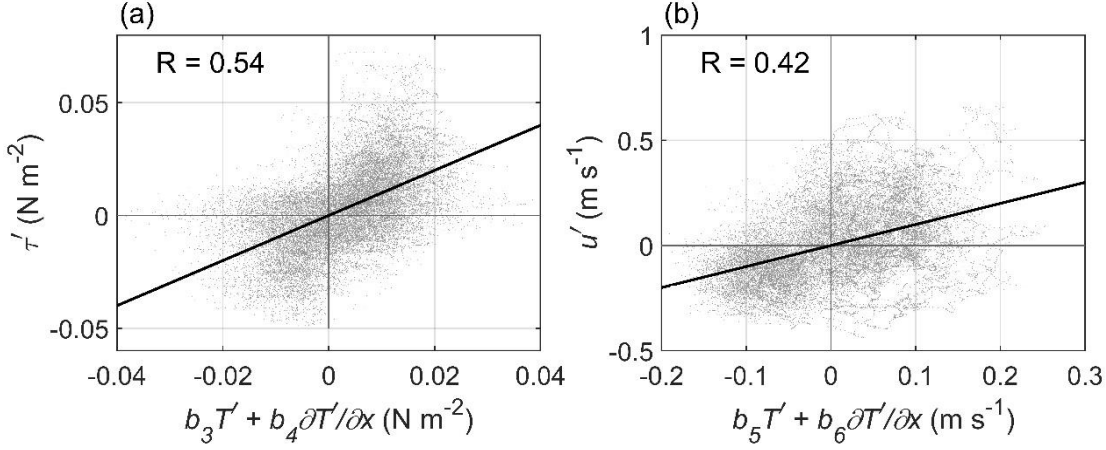

Fig. 8 The linear relations between the combination of SSTA and its gradient and the zonal wind stress anomaly (a), and the oceanic current anomaly (b). The data comes from the SODA version 3.4.2.

*b. Propagation Feature*

According to Eq. (23), an observer who moves with the zonal basic current (or an stationary observer due to zero zonal basic current) will see SSTA oscillation with no propagation. However, as Eq. (25) suggests, he or she may still observe wave propagation if the blocking effect of the boundary is quite weak and can be ignored. This works when a perturbation propagates toward a distant coastline. Besides, if the initial perturbation (the initial value) is only significant in a local region (e.g., in coastal regions) rather than the entire basin, the mutual interaction region between the westward propagating and eastward propagating waves will be near the initial region. This means that a region, if it is far away from the initial perturbation, is only influenced by the westward or the eastward propagating wave. Therefore, at the place far away from the initial field, Eq. (12) can be described by a simple westward propagating equation and a simple eastward propagating wave equation

$$\frac{\partial T'}{\partial t_1} - c\frac{\partial T'}{\partial x_1} = 0, \qquad (36)$$

$$\frac{\partial T'}{\partial t_1} + c\frac{\partial T'}{\partial x_1} = 0. \qquad (37)$$

Note that the net surface heating has been ignored here. Particularly, if the initial values only dominate near the western or eastern coastal region, Eq. (12) can be replaced with Eq. (36) or Eq. (37) to highlight the eastern or western propagating wave, that should be most significant in the central Pacific that is far away from the initial perturbation and far away from the coastlines. When the propagating wave is closing to the coastline, it may be absorbed or reflected by the coastline.

For some cases the eastern coastal SSTA may play roles, westward propagating waves may be observed. For example, Horel (1982) had pointed out that relatively warm SST appears in December-January along the coast of Peru and then spreads westward along the equator during the next several months. This propagating speed is around -0.5m s$^{-1}$ (Boucharel et al., 2013) and is basically in agreement with the wave propagation speed $\bar{u} - c = -0.35$m s$^{-1}$. Considering the fact that the zonal basic current in the eastern Pacific is generally larger than 0.1m s$^{-1}$ (Keenlyside, 2002), the theoretical propagation speed should be further closer to the observational propagation speed. For some cases the influence of the western Pacific SSTA may be more significant. Therefore, eastward propagating waves may be observed. For example, Simon Wang et al. (2015) found a systematic propagating pattern of SSTA has emerged between 100°E and 160°W, linking warm (cold) water in the western North Pacific to the development of El Niño (La Niña) in the central equatorial Pacific, for a duration of about 2–3 years. The corresponding propagation speed is about 0.1–0.16m s$^{-1}$, quite consistent with the eastward propagating speed $\bar{u} + c = 0.15$m s$^{-1}$.

The dispersion relation of the propagating waves can also be discussed by specifying the single wave solution, e.g., $T' \sim T_k \exp i(kx - \omega t)$, $Q' \sim Q_k \exp i(kx - \omega t)$ where $k$ is the wavenumber, $\omega$ is the frequency and $T_k$ and $Q_k$ are corresponding coefficients. Substituting them into Eq. (12) or Eq. (10), the dispersion relation is derived as

$$\omega'^2 - i\alpha_k \omega' - c^2 k^2 = 0, \qquad (38)$$

where $\omega' = \omega - \bar{u}k$ is the intrinsic frequency, $\alpha_k = Q_k/T_k$ is the ratio between the coefficients. Eq. (38) is a quadratic equation but with complex coefficients. It means the frequency will be complex, e.g., $\omega = \omega_r + i\omega_i$ where $\omega_r$ and $\omega_i$ are real values. Dividing it into real and imagery parts, it becomes

$$\omega_r = \bar{u}k \pm \sqrt{c^2 k^2 - \frac{1}{4}\alpha_k^2}$$
$$\omega_i = \frac{1}{2}\alpha_k$$
. (39)

The first equation in Eq. (39) demonstrates a propagating wave with energy dispersion. It is obvious that the net surface heating will modulate the wave frequency. The second equation in Eq. (39) means that the wave will develop to be unstable if $\alpha_k > 0$ but will decay if $\alpha_k < 0$. The net surface heating in the equatorial Pacific generally acts the thermodynamic damping (Jin et al., 2020). Therefore, we may generally set $\alpha_k < 0$. The dependence of $\alpha_k$ on the spatial scale (or the wavenumber) means waves with different spatial scales will have different decay rate. Besides, although $\alpha_k$ is generally smaller than zero, it may be larger than zero for specific wavenumbers in certain cases. This implies that the corresponding waves will be unstable.

If the net surface heating is replaced with a Newtonian cooling coefficient $\alpha$ for simplicity in Eq. (1), it is equivalent to set $\alpha_k = -\alpha < 0$ in Eq. (38). This results the waves, no matter what the wavenumbers are specified, will have a constant decline rate that equals the half the Newtonian cooling coefficient. Particularly, if neglecting the net surface heating force, we will further reduce Eq. (38) to a neutral wave with no energy dispersion

$$\omega = \omega_r = (\bar{u} \pm c)k.$$ (40)

## 6. Conclusions

A new model is built to capture the essence of the spatiotemporal variation of the ENSO phenomenon in this study. The model contains two variables, SSTA and TDA. The equation for SSTA considers the effects of the advection by the zonal basic current, the upwelling and the thermocline depth feedback processes, and the het surface heating. The equation for TDA takes the advection due to the zonal basic current and the divergence of the zonal current anomaly into consideration. To benefit theoretical analysis, the zonal current anomalies are firstly supposed to be proportional to the zonal wind stress anomalies based on the fact that the oceanic current in the mixed layer is driven by the wind stress. And the zonal wind stress anomalies are then hypothesized to be proportional to the gradients in SSTAs, rather than SSTAs themselves as in classic RO framework. Of course, a more precise hypothesis that emphasizes the linkage among the zonal wind stress anomalies, SSTAs, and gradients in SSTAs can also be proposed. With above two steps, a proportional relationship between the zonal current anomalies and the gradients in SSTAs is established to replace the terms that are associated with the zonal current anomalies with SSTA gradients. This helps to enclose the system to a classic wave equation for SSTA.

Given initial values and proper boundary conditions, the analytic solution is derived by applying cosine series expansion to separate the temporal and spatial variations of SSTA. The solution demonstrates that SSTA variations can be decomposed into the superposition of a series of spatial modes that oscillate with both natural frequencies and forced frequencies. The first spatial mode highlights SSTA contrast between the eastern and western Pacific. The warm (or cold) SSTAs appear in the eastern Pacific while cold (or warm) SSTAs are in the western Pacific. Ignoring the external forcing, it oscillates with a natural period of around 4.3 years. The spatial pattern of the first mode is similar to that in the EP El Niño events and its natural period is also consistent with the QQ mode. The second spatial mode emphasizes SSTA contrast between the central and eastern, western Pacific. The warm (or cold) SSTAs are located in the central Pacific while the cold (or warm) SSTAs are in the eastern and western Pacific. It oscillates with a natural period of around 2.3 year, that is the half of the first natural period. The spatial pattern of the second mode is similar to that in CP El Niño events and its natural period is quite consistent with the QB mode. Therefore, the first two modes can provide an explicit physical presentation of the spatial patterns of two types of ENSO and their temporal variations. Adding these two eigenmodes together with different weights, they can reproduce SSTA patterns in the strong El Niño events (e.g., twice the first plus the second) and SSTA patterns in the CP El Niño-like events (e.g., twice the second plus the first). The combined SSTA patterns show complex variations. For example, two La Niña events may alternatively appear after an El Niño event if observing from the NINO3.4 index. This may be enlightening for understanding the Triple Niña events that happened during in past 3 years (note that they had also appeared before) if considering more eigenmodes.

The net surface heating will exert its influence by introducing forced frequencies. Cosine series expansion for the net surface heating in SODA suggests that there exist many significant oscillation periods, such as 6 and 12 months for the first spatial mode, 6, 12, 30 and 114 months for the second spatial mode. It is obvious that these forcing periods will modulate natural periods to further complicate the temporal variations. Particularly, if the forcing period is close to the natural period, resonance will happen to amplify SSTA amplitude. Besides, the solution may also show certain propagation features if initial perturbations are limited in coastal regions and propagate toward a distant coastline so that the blocking effect of the coastline can be ignored. The dispersion relation suggests that the propagating wave of a specified wavenumber is energy dispersive and will decline with a rate that is determined by the net surface heating. The propagation feature may be applied in interpreting observed propagating phenomena.

Previous studies had focused on ENSO STPD (e.g., Timmermann et al., 2018; Jin, 2022). This new model decomposes the equatorial SSTA into a series of natural spatial modes, each of which has its natural oscillation that can be regarded as a classic RO model. It may also be developed by introducing nonlinear terms, stochastic terms and so on. Therefore, this new model can not only provide an explicit physical paradigm for understanding ENSO, but also a useful tool to be developed to address more challenges of ENSO STPD.